\documentclass{article}
\usepackage{amsmath,amssymb,cite,epsfig,graphicx,fancyhdr}
\begin{document}
\lhead{Non-hermitian Hamiltonians and $P_{IV}$ with real parameters}
\rhead{D. Bermudez, D. J. Fern\'andez C.}

\title{Non-hermitian Hamiltonians and \\ Painlev\'e IV equation with real parameters}
\author{David Berm\'udez\footnote{{\it email:} dbermudez@fis.cinvestav.mx}\,  
and David J. Fern\'andez C.\footnote{{\it email:} david@fis.cinvestav.mx} \\
{\sl Departamento de F\'{\i}sica, Cinvestav, A.P. 14-740, 07000 M\'exico D.F., Mexico}}

\date{}

\maketitle

\begin{abstract}
In this letter we will use higher-order supersymmetric quantum mechanics to obtain several families of complex solutions $g(x;a,b)$ of the Painlev\'e IV equation with real parameters $a,b$. We shall also study the algebraic structure, the eigenfunctions and the energy spectra of the corresponding non-hermitian Hamiltonians.\\

\end{abstract}

\section{Introduction}

Since its birth, supersymmetric quantum mechanics (SUSY QM) catalyzed the study of exactly solvable Hamiltonians and gave a new insight into the algebraic structure characterizing these systems. Historically, the essence of SUSY QM was developed first as Darboux transformation in mathematical physics \cite{MS91} and as factorization method in quantum mechanics \cite{IH51}.

On the other hand, there has been an increasing interest in the study of non-linear phenomena, which in many cases leads to the analysis of Painlev\'e equations \cite{VS93,Adl94}. Although these were discovered from strictly mathematical considerations, nowadays they are widely used to describe several physical phenomena \cite{AC92}. In particular, the Painlev\'e IV equation ($P_{IV}$) is relevant in fluid mechanics, non-linear optics and quantum gravity \cite{Win92}.

As it has been shown, there is a natural connection between quantum systems described by second-order polynomial Heisenberg algebras (PHA), whose Hamiltonians have the standard Schr\"odinger form and their differential ladder operators are of third order, and solutions $g(x;a,b)$ of $P_{IV}$ \cite{Adl94,ARS80,Fla80}. Moreover, these algebras can be realized by the $k$-th order SUSY partners $H_k$ of the harmonic oscillator Hamiltonian $H_0$, which leads to a simple method for generating solutions of $P_{IV}$, since the SUSY technique provides explicit expressions for the extremal states of $H_k$ and connecting formulae relating them with the corresponding $P_{IV}$ solutions \cite{ACIN00,FNN04,CFNN04,MN08}. It is worth to note, however, that the need to avoid singularities in the new potential $V_k(x)$ and the requirement for the Hamiltonian $H_k$ to be hermitian lead to some restrictions \cite{BF11}: (i) first of all, the relevant transformation function has to be real, which implies that the associated factorization energy is real; (ii) as a consequence, the spectrum of $H_k$ consists of two independent physical ladders, an infinite one departing from $E_0 = 1/2$ (the ground state energy of $H_0$) plus a finite one with $k$ equidistant levels, all of which have to be placed below $E_0$. Regarding $P_{IV}$, these two restrictions imply that non-singular real solutions $g(x;a,b)$ can be obtained for certain real parameters $a,b$.

From the point of view of spectral design, it would be important to overcome restriction (ii) so that some (or all) steps of the finite ladder could be placed above $E_0$. In this way we would be able to manipulate not just the lowest part of the spectrum (as we did previously \cite{FNN04,CFNN04,BF11}), but also the excited state levels, which would endow us with new tools for spectral design. In this article we are going to show that this can be achieved if one relaxes restriction (i) as well, which will force us to use complex transformation functions (see \cite{ACDI99}) and will lead to the generation of complex solutions to $P_{IV}$.

This letter is organized as follows: in Section 2 we shall present the general framework of SUSY QM and PHA. In the next Section we will generate the complex solutions to $P_{IV}$, specifically, we shall analyze the domain of the parameter space $(a,b)$ for which we obtain real or complex solutions; then, in Section 4 we will study the eigenfunctions and the energy spectra of the non-hermitian Hamiltonians. We shall present our conclusions in Section 5.

\section{General framework of SUSY QM and PHA}

In the $k$-th order SUSY QM one typically starts from a given solvable Hamiltonian
\begin{equation}
H_0  = -\frac12 \frac{d^2}{d x^2} + V_0(x),
\end{equation}
and generates a chain of standard (first-order) intertwining relations \cite{AIS93,MRO04,Fer10}
\begin{align}
H_j A_j^{+} & = A_j^{+} H_{j-1}, \quad  H_{j-1}A_j^{-} = A_j^{-}H_j, \\
H_j & = -\frac12 \frac{d^2}{d x^2} + V_j(x),\\
A_j^{\pm} &= \frac{1}{\sqrt{2}}\left[\mp \frac{d}{d x} + \alpha_j(x,\epsilon_j)\right], \quad j = 1,\dots,k. 
\end{align}
Hence, the following equations must be satisfied
\begin{align}
& \alpha_j'(x,\epsilon_j) + \alpha_j^2(x,\epsilon_j) = 2[V_{j-1}(x) - \epsilon_j],  \label{rei} \\
& V_{j}(x) = V_{j-1}(x) - \alpha_j'(x,\epsilon_j). \label{npi}
\end{align}
We are interested in the final Riccati solution $\alpha_{k}(x,\epsilon_{k})$, which turns out to be determined, either by $k$ solutions $\alpha_1(x,\epsilon_j)$ of the initial Riccati equation
\begin{equation}
\alpha_1'(x,\epsilon_j) + \alpha_1^2(x,\epsilon_j) = 2 [V_0(x) - \epsilon_j], \quad j=1,\dots,k,
\end{equation}
or by $k$ solutions $u_j$ of the associated Schr\"odinger equation
\begin{equation}
H_0 u_j = - \frac12 u_j'' + V_0(x)u_j = \epsilon_j u_j, \quad j=1,\dots,k, \label{usch}
\end{equation}
with $\alpha_1(x,\epsilon_j) = u_j'/u_j$.

Thus, there is a pair of $k$-th order operators interwining the initial $H_0$ and the final Hamiltonians $H_k$, namely,
\begin{equation}
H_k B_k^{+} = B_k^{+} H_0, \quad H_0 B_k^{-} = B_k^{-} H_k,
\end{equation}
where
\begin{equation}
B_k^{+} = A_k^{+}\dots A_1^{+}, \quad B_k^{-} = A_1^{-}\dots A_k^{-}.
\end{equation}

The normalized eigenfunctions $\psi_n^{(k)}$ of $H_k$, associated to the eigenvalues $E_n$, are proportional to the action of $B_k^{+}$ onto the corresponding ones of $H_0$ ($\psi_n$, $n=0,1,\dots$). Moreover, there are $k$ additional eigenstates $\psi_{\epsilon_j}^{(k)}$ associated to the eigenvalues $\epsilon_j$ ($j=1,\dots ,k$), which are simultaneously annihilated by $B_k^{-}$. Their corresponding explicit expressions are given by \cite{BF11,FH99}:
\begin{align}
\psi_n^{(k)} = \frac{B_k^{+}\psi_n}{[(E_n-\epsilon_1)\dots (E_n-\epsilon_k)]^{1/2}}, & \quad E_n, \label{psin}\\
\psi_{\epsilon_j}^{(k)} \propto \frac{W(u_1,\dots , u_{j-1},u_{j+1},\dots , u_k)}{W(u_1,\dots , u_k)}, & \quad \epsilon_j. \label{psie}
\end{align}
Let us note that, in this formalism the obvious restriction $\epsilon_j < E_0=1/2$ naturally arises since we want to avoid singularities in $V_k(x)$.

On the other hand, a $m$-th order PHA is a deformation of the Heisenberg-Weyl algebra of kind \cite{CFNN04,FH99}:
\begin{align}
[H,L^\pm] 		&= \pm L^\pm , \\
[L^-,L^+] 			& \equiv Q_{m+1}(H+1) - Q_{m+1}(H) = P_m(H) , \\
Q_{m+1}(H) 	&= L^+ L^- = \prod\limits_{i=1}^{m+1} \left(H - \mathcal{E}_i\right) ,
\end{align}
where $Q_{m+1}(x)$ is a $(m+1)$-th order polynomial in $x$, which implies that $P_m(x)$ is a polynomial of order $m$ in $x$ and $\mathcal{E}_i$ are the zeros of $Q_{m+1}(H)$, which correspond to the energies associated to the extremal states of $H$.

Now, let us take a look at the differential representation of the second-order PHA ($m=2$). Suppose that $L^+$ is a third-order differential ladder operator, chosen by simplicity as:
\begin{align}
L^+   &= L_1^+ L_2^+ , \\
L_1^+ &= \frac{1}{\sqrt{2}}\left[-\frac{d}{d x} + f(x) \right], \\
L_2^+ &= \frac12\left[ \frac{d^2}{d x^2} + g(x)\frac{d}{d x} +
h(x)\right].
\end{align}
These operators satisfy the following intertwining relationships:
\begin{align}
HL_1^+ & = L_1^+ (H_{\rm a} + 1), \quad H_{\rm a} L_2^+ = L_2^+ H,\\
 \Rightarrow \quad & [H,L^+] = L^+,
\end{align}
where $H_{\rm a}$ is an auxiliary Schr\"odinger Hamiltonian. Using the standard equations for the first and second-order SUSY QM and decoupling the resulting system gives rise to
\begin{align}
& f = x + g, \label{fdependg} \\
& h = - x^2 + \frac{g'}{2} - \frac{g^2}{2} - 2xg + a, \\
& V = \frac{x^2}2 - \frac{g'}2 + \frac{g^2}2 + x g + \mathcal{E}_1 -
\frac12 , \label{Vpivs}
\end{align}
where
\begin{equation}
g'' = \frac{g'^2}{2g} + \frac{3}{2} g^3 + 4xg^2 + 2\left(x^2 - a \right) g + \frac{b}{g}.
\end{equation}
Note that this is the Painlev\'e IV equation ($P_{IV}$) with parameters
\begin{equation}
a =\mathcal{E}_2 + \mathcal{E}_3-2\mathcal{E}_1 -1,\quad b = - 2(\mathcal{E}_2 - \mathcal{E}_3)^2.\label{abe}
\end{equation}
Hence, if the three quantities $\mathcal{E}_i$ are real, we will obtain real parameters $a,b$ for the corresponding $P_{IV}$.

Let us note that the potential of equation (\ref{Vpivs}) contains a harmonic oscillator term. In addition, three terms can be identified ($-g'/2 + g^2/2 + xg$) which in general lead to an anharmonicity in the potential and are completely determined by the solution to $P_{IV}$. As a consequence, one could say that the solution $g$ to $P_{IV}$ is the main responsible for the spectral differences which the Hamiltonian $H$ could have with respect to the harmonic oscillator (compare with \cite{DEK94}).

\section{Complex solutions to $P_{IV}$ with real parameters}

It is well known that the first-order SUSY partner Hamiltonians of the harmonic oscillator are naturally described by second-order PHA, which are connected with $P_{IV}$, as we have shown in the previous section. Furthermore, there is a theorem stating the conditions for the hermitian higher-order SUSY partners Hamiltonians of the harmonic oscillator to have this kind of algebras (see \cite{BF11}). The main requirement is that the $k$ Schr\"odinger seed solutions have to be connected in the way
\begin{align}
u_j=(a^{-})^{j-1}u_1,& \quad \label{us}\\
\epsilon_j=\epsilon_1-(j-1), & \quad j=1,\dots , k,
\end{align}
where $a^{-}$ is the standard annihilation operator of $H_0$ so that the only free seed $u_1$ has to be a real solution of Eq.~\eqref{usch} without zeros, associated to a real factorization energy $\epsilon_1$ such that $\epsilon_1<E_0=1/2$.

In this work we intend to overcome this restriction, although if we use the formalism as in \cite{BF11} with $\epsilon_1 > E_0$, we would obtain only singular SUSY transformations. In order to avoid this we will instead employ complex SUSY transformations. The simplest way to implement them is to use a complex linear combination of the two standard linearly independent real solutions which, up to an unessential factor, leads to the following complex solutions depending on a complex constant $\lambda + i \kappa$ ($\lambda, \kappa \in \mathbb{R}$) \cite{ACDI99}:
\begin{equation}
u(x;\epsilon ) = e^{-x^2/2}\left[ {}_1F_1\left(\frac{1-2\epsilon}{4},\frac12;x^2\right)
 + x(\lambda + i\kappa)\, {}_1F_1\left(\frac{3-2\epsilon}{4},\frac32;x^2\right)\right], \label{u1}
\end{equation}
where $_1F_1$ is the confluent hypergeometric (Kummer) function. The known results for the real case \cite{JR98} are obtained by making $\kappa=0$ and expressing $\lambda$ as
\begin{equation}
\lambda= 2 \nu\frac{\Gamma(\frac{3 - 2\epsilon}{4})}{\Gamma(\frac{1-2\epsilon}{4})}, \label{nu}
\end{equation}
with $\nu \in \mathbb{R}$.

Hence, through this formalism we will obtain the $k$-th order SUSY partner potential $V_k(x)$ of the harmonic oscillator and the corresponding $P_{IV}$ solution $g(x;\epsilon_1)$, both of which are complex, in the way
\begin{align}
V_k(x) &= \frac{x^2}2 - \{\ln [W(u_1,\dots,u_k)]\}'' , \\
g(x;\epsilon_1) &= - x - \{\ln[\psi_{\mathcal{E}_1}(x)]\}'. \label{solg}
\end{align}
Note that the extremal states of $H_{k}$ and their corresponding energies are given by
\begin{align}
\psi_{\mathcal{E}_1} \propto \frac{W(u_1,\dots,u_{k-1})}{W(u_1,\dots,u_k)}, & \quad \mathcal{E}_1 = \epsilon_k = \epsilon_1 - (k - 1), \label{edo1}\\
\psi_{\mathcal{E}_2} \propto B_k^+ e^{-x^2/2}, & \quad \mathcal{E}_2 = \frac{1}{2}, \label{edo2}\\
\psi_{\mathcal{E}_3} \propto B_k^+ a^{+} u_1, & \quad \mathcal{E}_3 = \epsilon_1 + 1. \label{edo3}
\end{align}
Recall that all the $u_j$ satisfy Eq.~\eqref{us} and $u_1$ corresponds to the general solution given in Eq.~\eqref{u1}.

\begin{figure}
\begin{center}
\includegraphics[scale=0.38]{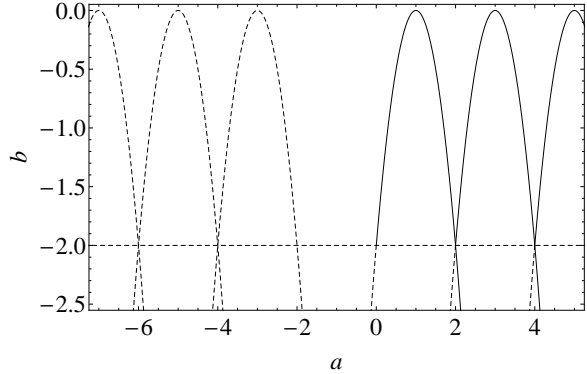}
\end{center}
\vspace{-5mm}
\caption{Parameter space $(a,b)$ of the $P_{IV}$ solutions. The curves represent the solution subspace for non-singular real or complex (solid curves) and only complex (dashed curves) solutions.} \label{parameterspace}
\end{figure}

For $k=1$, the first-order SUSY transformation  and Eq.~\eqref{solg} lead to what is known as \emph{one-parameter solutions} to $P_{IV}$, due to the restrictions imposed by Eq.~\eqref{abe} onto the two parameters $a,b$ of $P_{IV}$ which makes them both depend on $\epsilon_1$ \cite{BCH95}. For this reason, this family of solutions cannot be found in any point of the parameter space $(a,b)$, but only in the subspace defined by the curve $\{\left( a(\epsilon_1), b(\epsilon_1)\right),\ \epsilon_1 \in \mathbb{R}\}$ consistent with Eqs.~\eqref{abe}. Then, by increasing the order of the SUSY transformation to an arbitrary integer $k$, we will expand this subspace to obtain $k$ different families of one-parameter solutions. This procedure is analogous to iterated auto-B\"acklund transformations \cite{RS82}. Also note that by making cyclic permutations of the indices of the three energies $\mathcal{E}_i$ and the corresponding extremal states of Eqs.~(\ref{edo1}-\ref{edo3}), we expand the solution families to three different sets, defined by
\begin{align}
a_{1}=-\epsilon_1 + 2k -\frac{3}{2}, \quad & b_{1}=-2\left(\epsilon_1+\frac{1}{2}\right)^{2}, \label{ab1}\\
a_2= 2\epsilon_1 -k, \quad & b_2=-2k^2, \\
a_3=-\epsilon_1-k-\frac{3}{2}, \quad & b_3=-2\left(\epsilon_1 - k +\frac{1}{2}\right)^2,
\end{align}
where we have added an index corresponding to the extremal state (Eqs.~(\ref{edo1}-\ref{edo3})) taken to build up the $P_{IV}$ solution in Eq.~\eqref{solg}. The first pair, $a_1,b_1$, can provide non-singular real or complex solutions, while the second and third ones can give just non-singular complex solutions, due to singularities in the real case. A part of the non-singular solution subspace for both real and complex cases is shown in Fig.~\ref{parameterspace}. One can check that those points which belong to two different sets have associated the same $P_{IV}$ solutions.

\begin{figure}
\begin{center}
\includegraphics[scale=0.38]{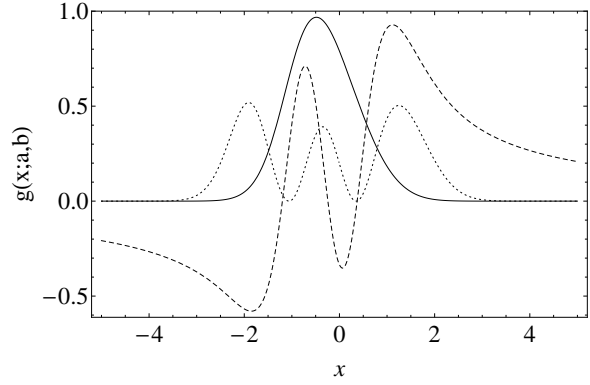}
\end{center}
\vspace{-5mm}
\caption{Some real solutions to $P_{IV}$, corresponding to $a_1=1$, $b_1=0$ ($k=1$, $\epsilon_1=-1/2$, $\nu=0.7$) (solid curve), $a_1=4$, $b_1=-2$ ($k=2$, $\epsilon_1=-3/2$, $\nu=0.5$) (dashed curve), and  $a_1=7$, $b_1=-8$ ($k=3$, $\epsilon_1=-1/2$, $\nu=0.3$) (dotted curve).} \label{greal}
\end{figure}

In turn, let us analyze some of the $P_{IV}$ solutions obtained by this method. The real solutions arise by taking  $\kappa=0$, and expressing $\lambda$ as in Eq. \eqref{nu} with $\epsilon_1<1/2$. They can be classified into three relevant solution hierarchies, namely, confluent hypergeometric, complementary error and rational hierarchies. Let us note that the same set of real solutions to $P_{IV}$ can be obtained through inverse scattering techniques \cite{AC92} (compare the solutions of \cite{BCH95} with those of \cite{BF11}). In Fig.~\ref{greal}, three real solutions to $P_{IV}$ are presented, which belong to the complementary error hierarchy.

\begin{figure}
\begin{center}
\includegraphics[scale=0.46]{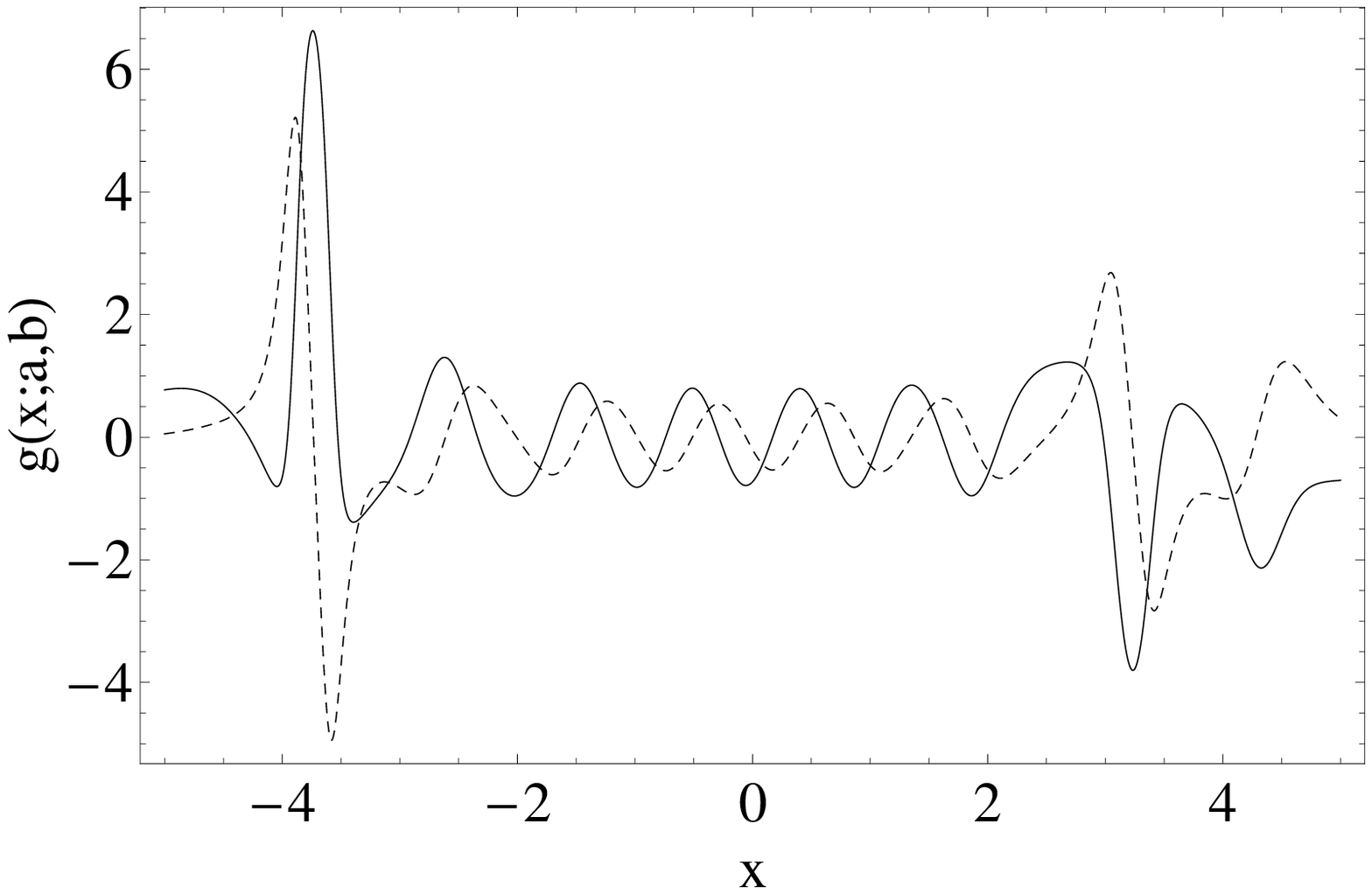}\hspace{5mm}
\includegraphics[scale=0.37]{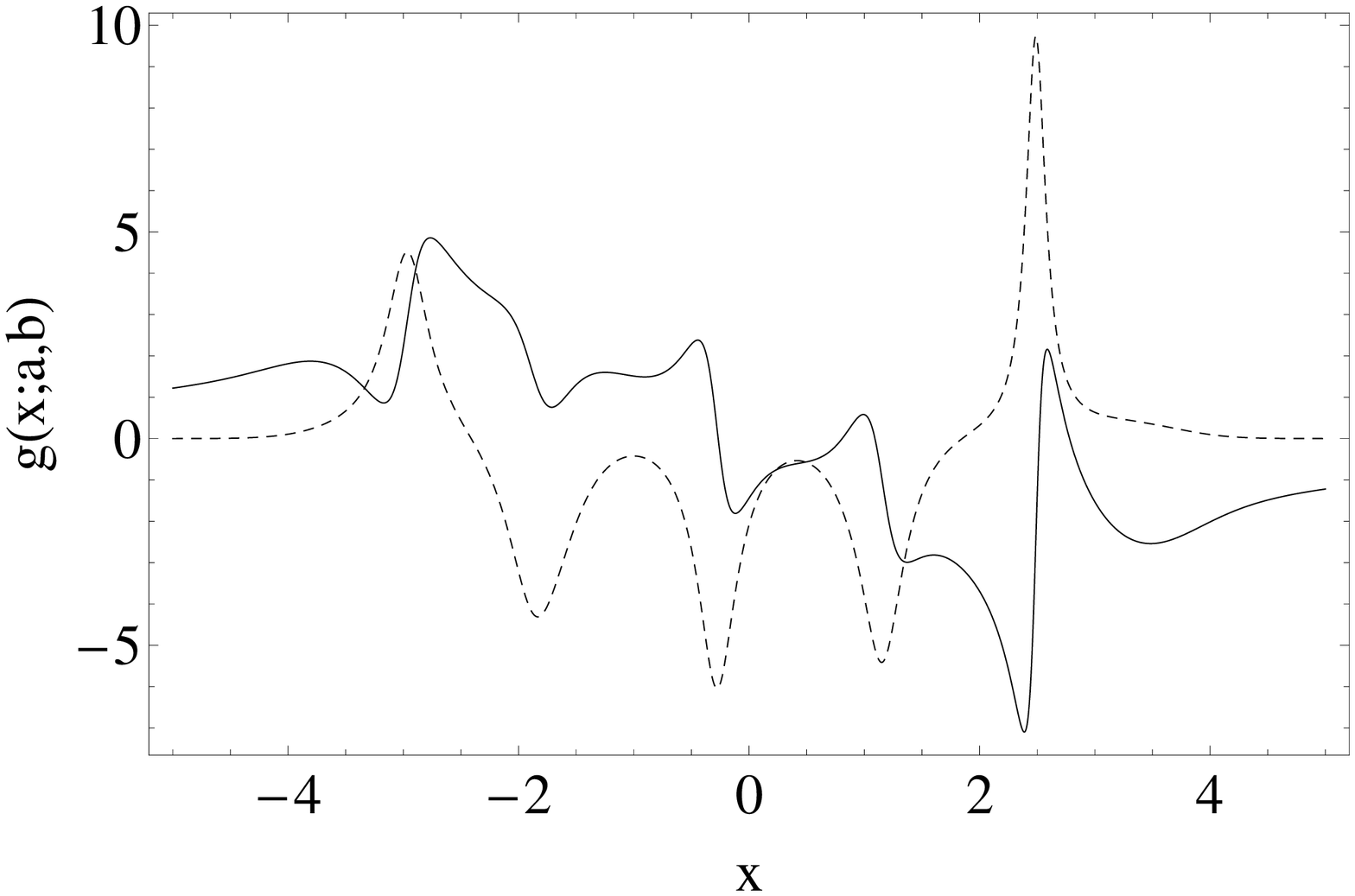}
\end{center}
\vspace{-5mm}
\caption{Real (solid curve) and imaginary (dashed curve) parts of some complex solutions to $P_{IV}$. The upper plot corresponds to $a_2=12$, $b_2=-8$ ($k=2$, $\epsilon_1=7$, $\lambda=\kappa=1$) and the lower one to $a_3=-5$, $b_3=-8$ ($k=1$, $\epsilon_1=5/2$, $\lambda=\kappa=1$).} \label{gcomplex}
\end{figure}

Next, we study the complex solutions subspace, i.e. we allow now that $\epsilon_1 \geq 1/2$. The real and imaginary parts of the complex solutions $g(x;a,b)$ for two particular choices of real parameters $a,b$, which belong to different solution sets, are plotted in Fig.~\ref{gcomplex}.

\begin{figure}
\begin{center}
\includegraphics[scale=0.3]{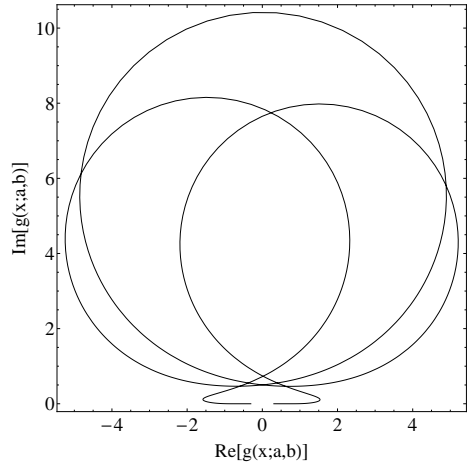}
\end{center}
\vspace{-5mm}
\caption{Parametric plot of the real and imaginary parts of $g(x;a,b)$ for  $a_1=-6$, $b_1=-2$ ($k=1$, $\epsilon_1=5/2$, $\lambda=1$, $\kappa=5$) and $|x| \leq 10$. For bigger values of $x$, the curve approaches the origin in both sides.} \label{complexpara}
\end{figure}

Note that, in general, $\psi_{\mathcal{E}_i}\neq 0\ \forall\ x \in \mathbb{R}$ , i.e., the solutions $g(x;a,b)$ are not singular. Moreover, both real and imaginary parts have an asymptotic null behaviour ($g\rightarrow 0$ as $|x|\rightarrow \infty$). This property becomes evident in Fig.~\ref{gcomplex}, as well as in the parametric plot of the real and imaginary parts of $g(x;a,b)$ of Fig.~\ref{complexpara}.

\section{Non-hermitian Hamiltonians}

Let us analize the Hamiltonian $H_k$ obtained by the complex SUSY transformation. Note that the real case, which leads to hermitian Hamiltonians, has been studied previously \cite{AIS93,Mie84}, allowing to obtain some criteria related to the structure of the associated energy spectrum $\text{Sp}(H_k)$, the number of zeros of the eigenfunctions of $H_k$, and the way in which they are connected by the third-order ladder operators $L^{\pm}$. This action is in agreement with the fact that $\text{Sp}(H_k)$ consists of an infinite ladder plus a finite one: there are two extremal states (both annihilated by $L^{-}$) from which the two ladders start, one associated to $\epsilon_k$ and the other one to $E_0=1/2$; since the ladder starting from $\epsilon_k$ ends at $\epsilon_1$, the eigenfunction associated to $\epsilon_1$ is annihilated by $L^{+}$. The actions of $L^{\pm}$ onto any other eigenstate of $H_k$ are non-null, and connect only the eigenstates belonging to the same ladder.

As far as we know, complex SUSY transformations with real factorization energies were used for the first time by Andrianov et al. to obtain non-hermitian Hamiltonians with real spectra \cite{ACDI99}. These topics have been of great interest in the context of both parity-time (PT) symmetric Hamiltonians (see Bender et al. \cite{BB98}) and pseudo-hermitian Hamiltonians (see Mostazafadeh et al. \cite{MB04}). Next, we will examine the structure of non-hermitian SUSY generated Hamiltonians $H_k$.

\begin{figure}
\begin{center}
\includegraphics[scale=0.1]{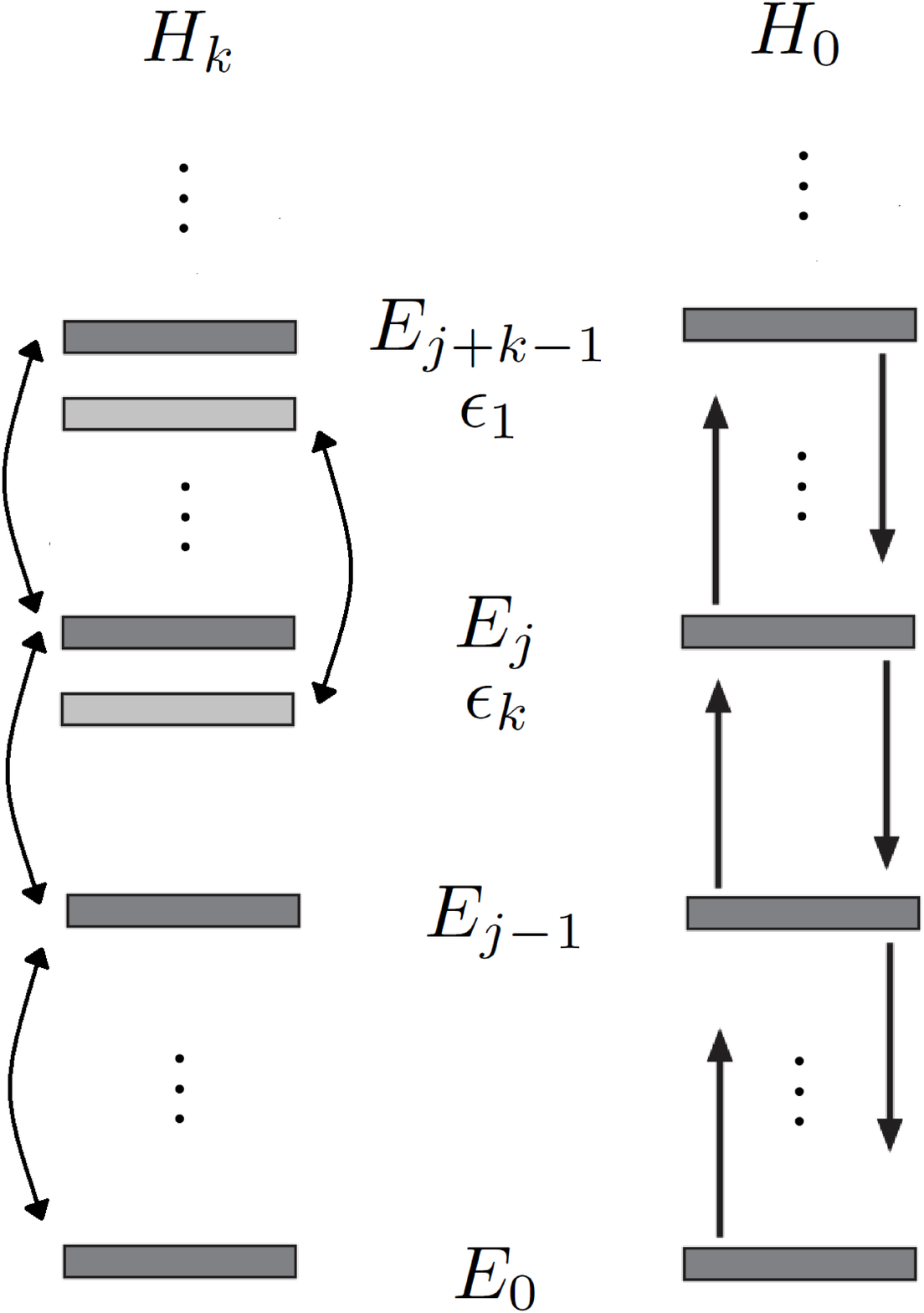}
\end{center}
\vspace{-5mm}
\caption{Spectrum of the SUSY partner Hamiltonians $H_0$ (right) and $H_k$ (left) for $\epsilon_1>1/2$, $\epsilon_1\neq E_j$; Sp($H_k$) still contains one finite and one infinite ladders. The dark bars represent the original and mapped eigenstates of $H_0$ and $H_k$, while the light ones the $k$ new levels of $H_k$ introduced by the $k$-th order SUSY transformation. All of them have associated square-integrable eigenfunctions.} \label{espectros}
\end{figure}

First of all, the new Hamiltonians necessarily have complex eigenfunctions, although the associated eigenvalues are still real. In previous works, the factorization energy associated to the real transformation function $u_1$ was bounded, $\epsilon_1<E_0=1/2$. In this paper we are using complex transformation functions to be able to overcome this restriction and yet obtain non-singular solutions. This naturally leads to complex solutions to $P_{IV}$ generated through factorization energies which could be placed now above $E_0$. The resulting spectra for the non-hermitian Hamiltonians $H_k$ obey the same criteria as the real case, namely, they are composed of an infinite ladder plus a finite one, which now could be placed, either fully or partially, above $E_0$. The eigenfunctions associated to the energy levels of the original harmonic oscillator are given by Eq.~\eqref{psin} and the ones associated to the new energy levels by Eq.~\eqref{psie}, all of them square-integrable. A diagram of the described spectrum is shown in Fig. \ref{espectros}.

\begin{figure}[h]
\begin{center}
\includegraphics[scale=0.37]{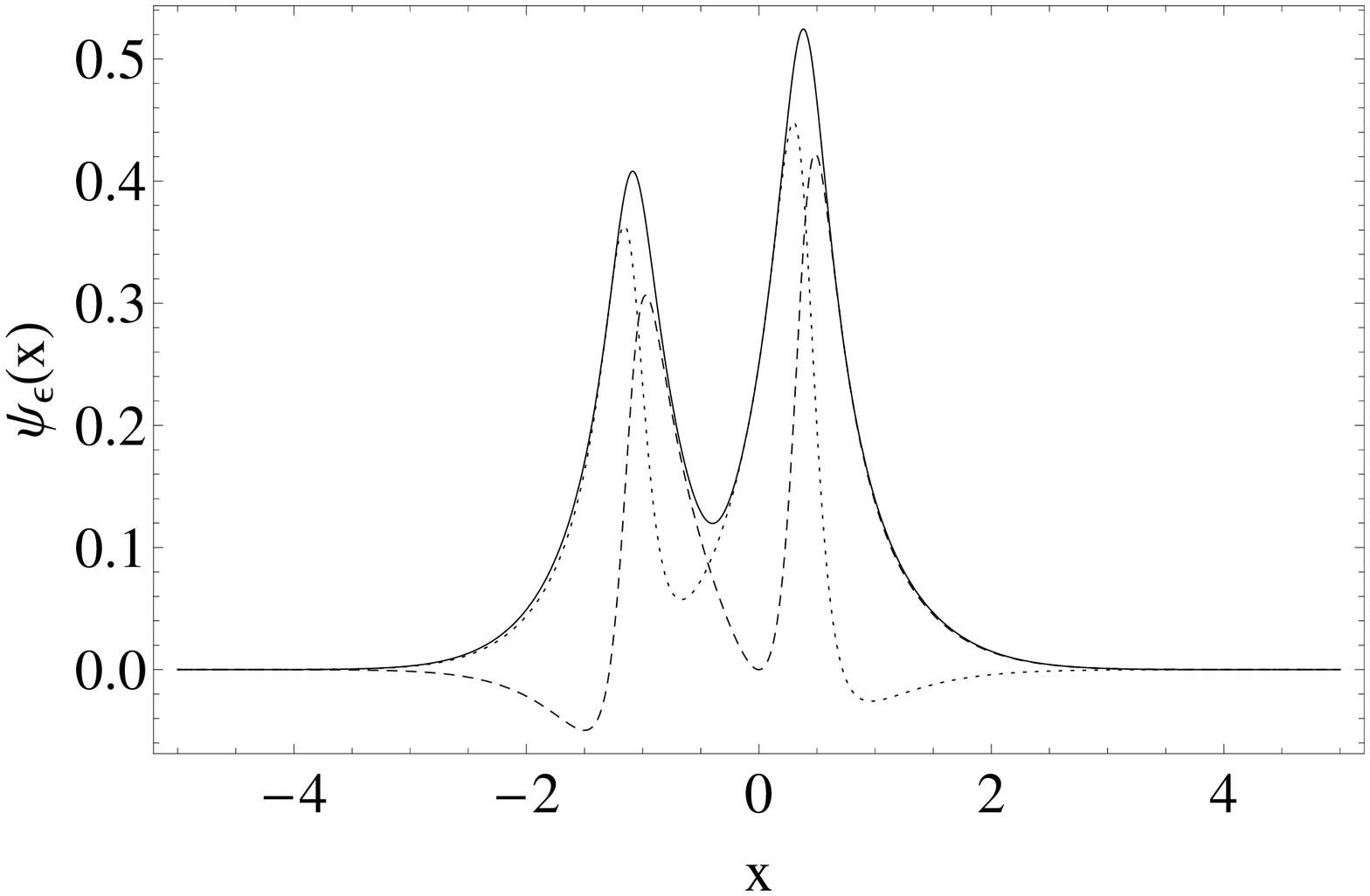}\\
\hspace{-4mm}\includegraphics[scale=0.393]{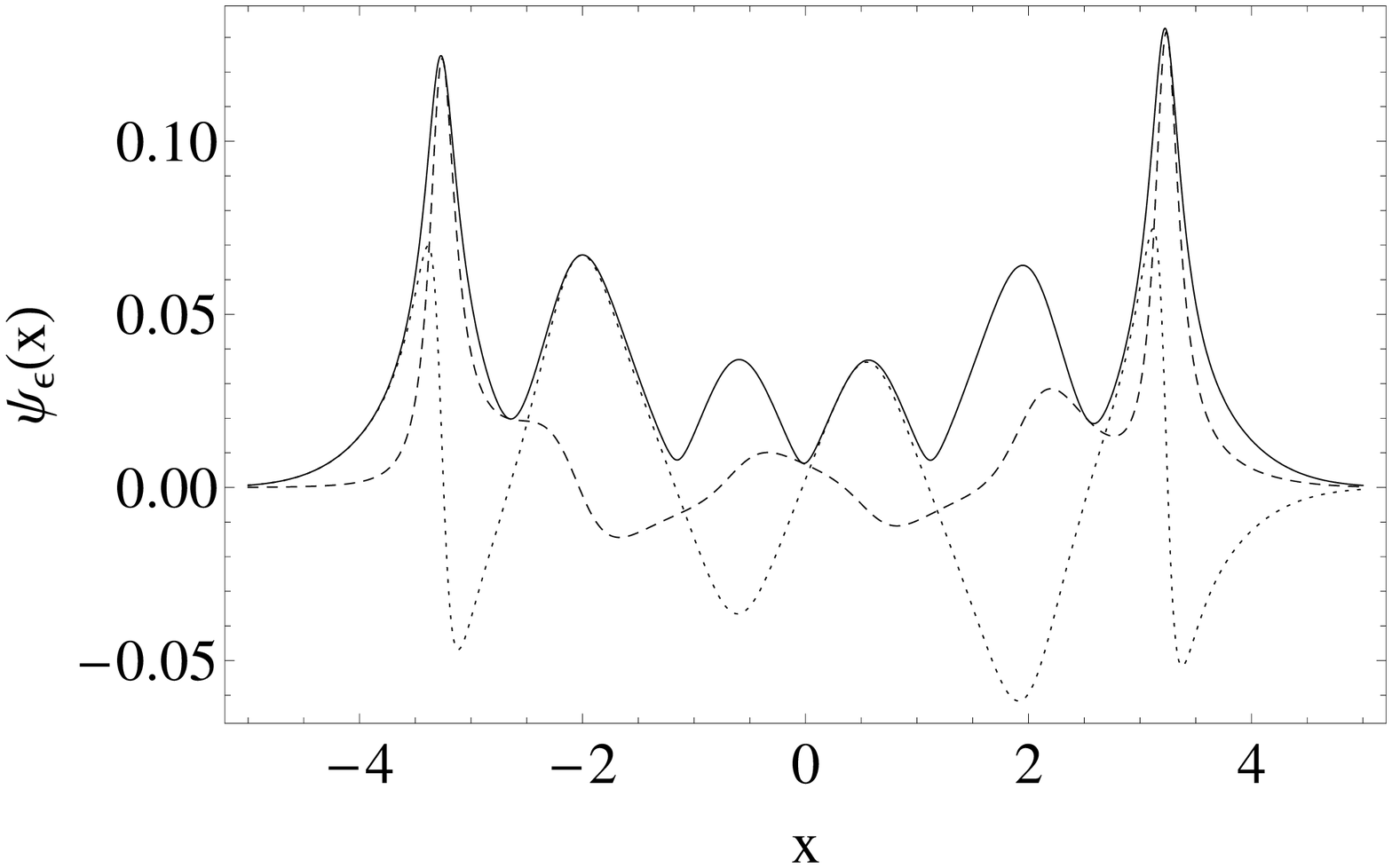}
\end{center}
\vspace{-5mm}
\caption{Plot of the absolute value, the real, and the imaginary parts (solid, dashed and dotted lines, respectively) of the eigenfunction $\psi_{\mathcal{E}_1}(x)$ given by Eq.~\eqref{edo1} for the values $k=2$, $\epsilon_1=-1$, $\lambda=1$, $\kappa=1/2$ (up) and $k=2$, $\epsilon_1=4$, $\lambda=1$, $\kappa=6$ (down).} \label{waves}
\end{figure}

The extremal states of the SUSY generated Hamiltonian $H_k$ are given by Eqs.~(\ref{edo1}-\ref{edo3}). These are non-singular complex eigenfunctions of $H_k$ and, from their asymptotic behaviour, we conclude that those given by Eqs.~(\ref{edo1},\ref{edo2}) are square-integrable. Note that in this case the oscillatory theorem does not hold anymore, neither for the real nor for the imaginary parts, although a related node structure emerges. The absolute value and the real and imaginary parts of $\psi_{\mathcal{E}_1}(x)$ for two particular cases are shown in Fig. \ref{waves}.

\begin{figure}
\begin{center}
\includegraphics[scale=0.1]{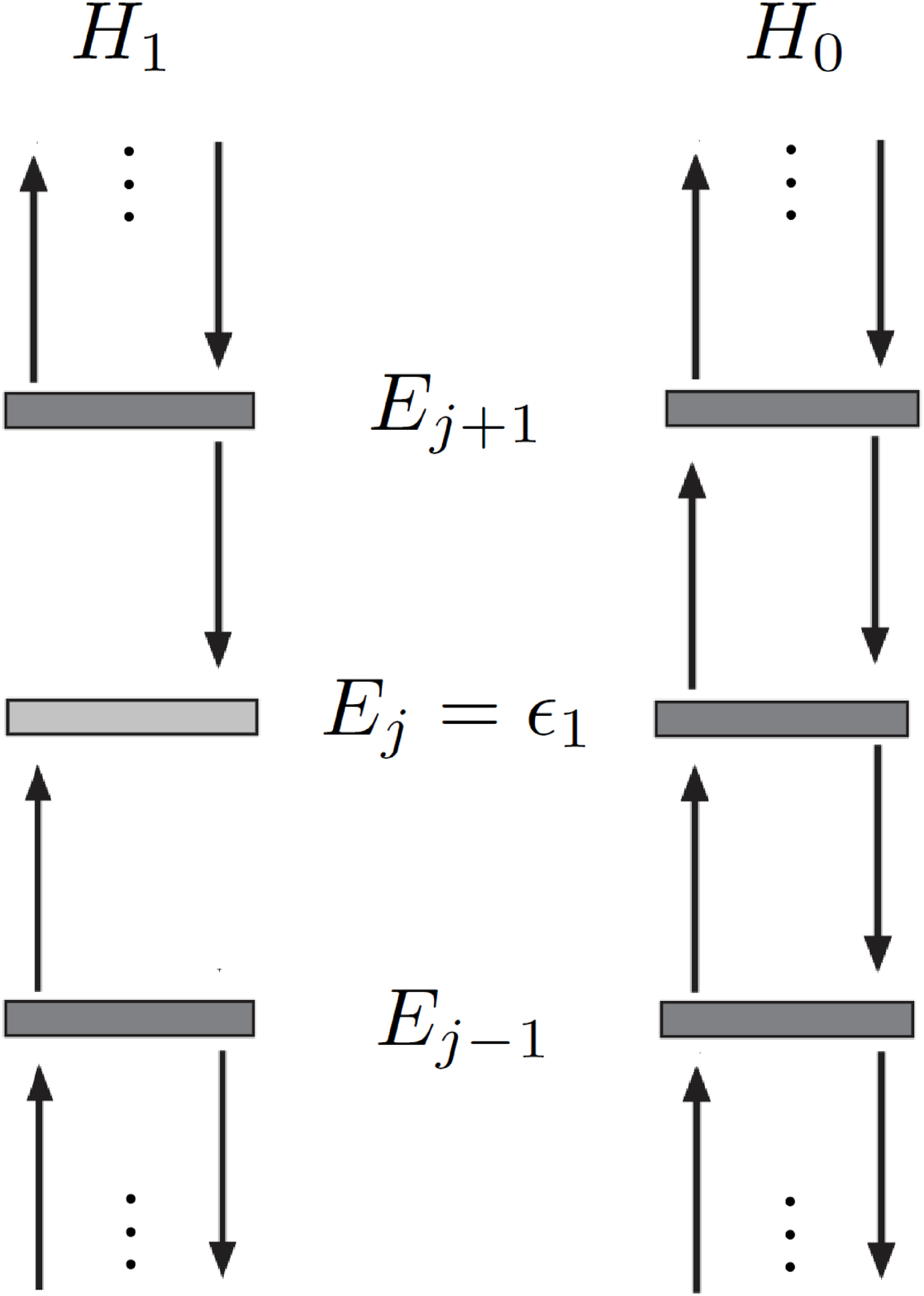}
\end{center}
\vspace{-5mm}
\caption{Spectra of the harmonic oscillator Hamiltonian $H_0$ and of its first-order SUSY partner $H_1$ when we use the factorization energy $\epsilon_1=E_j \in \text{Sp}(H_0)$. The level $E_j$ of $H_1$ is connected with its adjacent ones just in one way.} \label{1susy}
\end{figure}

On the other hand, complex transformations for $\epsilon_1=E_j$ are worth of a detailed study, namely, when the factorization energy $\epsilon_1$ belongs to the spectrum of the original harmonic oscillator Hamiltonian. For instance, let us consider a first-order SUSY transformation with $\epsilon_1=E_j$ and $u_1$ given by Eq.~\eqref{u1}, i.e., $u_1$ is a complex linear combination of the eigenfunction $\psi_j$ of $H_0$ and the other linearly independent solution of the Schr\"odinger equation. It is straightforward to see that the action of the ladder operator $L^{-}=A_1^{+}a^{-}A_1^{-}$ is given by
\begin{align}
E_{l},&\quad L^{-}(A_1^{+}\psi_{l})  \propto A_1^{+}\psi_{l-1},\\
E_j, &\quad L^{-}(A_1^{+}\psi_{j})  = 0,\\
E_{0}, &\quad L^{-}(A_1^{+}\psi_{0})  = 0,
\end{align}
where $l \neq j$, $l \neq 0$, the shown energies correspond to the departure state, and we have used that $A_1^{+}\psi_j \propto 1/u_1 $. For $L^{+}=A_1^{+}a^{+}A_1^{-}$ we have
\begin{align}
E_{l},&\quad L^{+}(A_1^{+}\psi_{l})  \propto \psi_{l+1},\\
E_j, &\quad L^{+}(A_1^{+}\psi_{j})  =0,
\end{align}
which does not match with the established criteria for the non-singular real and complex cases with $\epsilon_1 \neq E_j$ since now it turns out that:
\begin{align}
E_{j+1},&\quad L^{-}(A_1^{+}\psi_{j+1})  \propto A_1^{+}\psi_{j} \propto \frac{1}{u_1}\neq 0,\\
E_{j-1},&\quad L^{+}(A_1^{+}\psi_{j-1})   \propto A_1^{+}\psi_{j} \propto \frac{1}{u_1}\neq 0.
\end{align}
The resulting Hamiltonian is isospectral to the harmonic oscillator but with a special algebraic structure because now one state (the one associated to $E_j$) is connected just in one way with the adjacent ones (associated to $E_j \pm 1$). A diagram representing this structure is shown in Fig.~\ref{1susy}. We are currently studying the $k$-th order case and expect to find the new criteria which will be valid for these special transformations.

\section{Conclusions}

In this letter, based on PHA and higher-order SUSY QM, we have introduced a method to obtain real and complex solutions $g(x;a,b)$ of the $P_{IV}$ with real parameters $a,b$. We have studied the properties of the resulting solutions, including the analysis of the subspace of the parameter space $(a,b)$ were non-singular real or complex solutions can be found. In addition, we have analyzed the algebras, the eigenfunctions and the spectra of the non-hermitian SUSY generated Hamiltonians.

Further investigation on the description of the analytic structure of the complex solutions and on the way in which the ladder operators $L^{\pm}$ map between different eigenstates and their corresponding energy levels is needed. Besides, we are looking for extensions of this technique to obtain $P_{IV}$ solutions associated to complex parameters $a,b$.

\section*{Acknowledgement}

The authors acknowledge the support of Conacyt.

\end{document}